\documentclass{pasj00}

\begin{document}
\SetRunningHead{T.Yamamoto et al.}{Discovery of cyclotron resonance feature from GX 304$-$1}

\title{ Discovery of a Cyclotron Resonance Feature in the X-ray Spectrum
  of GX 304$-$1 with RXTE and Suzaku during Outbursts Detected \\
by MAXI in 2010 }

\author{%
  Takayuki \textsc{Yamamoto},\altaffilmark{1,2}
  Mutsumi \textsc{Sugizaki},\altaffilmark{2}
  Tatehiro \textsc{Mihara},\altaffilmark{2}
  Motoki \textsc{Nakajima},\altaffilmark{3}
  Kazutaka \textsc{Yamaoka},\altaffilmark{4}
  Masaru \textsc{Matsuoka},\altaffilmark{2}
  Mikio \textsc{Morii} \altaffilmark{5},
  and
  Kazuo \textsc{Makishima},\altaffilmark{2,6}
}

\altaffiltext{1}{Department of Physics, Nihon University, 
  1-8-14 Surugadai, Chiyoda, Tokyo 101-8308, Japan}
\altaffiltext{2}{MAXI team, RIKEN, 2-1 Hirosawa, Wako, Saitama 351-0198, Japan}
\email{tyamamot@crab.riken.jp}
\altaffiltext{3}{School of Dentistry at Matsudo, Nihon University, 
  2-870-1 Sakaecho-nishi, Matsudo, Chiba 101-8308, Japan}
\altaffiltext{4}{Department of Physics and Mathematics, Aoyama Gakuin University, 
  5-10-1 Fuchinobe, Chuo, Sagamihara, Kanagawa 252-5258, Japan}
\altaffiltext{5}{Department of Physics, Tokyo Institute of Technology,
  2-12-1 Ookayama, Meguro-ku, Tokyo 152-8551, Japan}
\altaffiltext{6}{Department of Physics, The University of Tokyo, 
  7-3-1 Hongo, Bunkyo, Tokyo 113-0033, Japan}


\maketitle

\begin{abstract}

  We report the discovery of a cyclotron resonance scattering feature (CRSF)
  in the X-ray spectrum of GX 304$-$1, obtained by RXTE and Suzaku during
  major outbursts detected by MAXI in 2010.  The
  peak intensity in August reached 600 mCrab in the 2--20
  keV band, which is the highest ever observed from this source.
  The RXTE observations on more than twenty
  occasions and one Suzaku observation revealed a spectral absorption
  feature at around 54 keV, 
  which is the first
  CRSF detection from this source. The estimated strength of surface magnetic
  field, $4.7\times 10^{12}$ G, is one of the highest among binary X-ray
  pulsars from which CRSFs have ever been detected.  The RXTE
  spectra taken during the August outburst also suggest that the CRSF
  energy changed over 50--54 keV, possibly in a positive
  correlation with the X-ray flux.  The behavior is qualitatively
  similar to that observed from Her X-1 on long time scales, or from A 0535$+$26, but different from
  the negative correlation observed from 4U~0115+63 and X~0331+53.

\end{abstract}

\section{Introduction}
\label{sec:intro}

The magnetic field strength of neutron stars is one of the
important parameters related to their fundamental physics.  The
surface magnetic field of accreting X-ray pulsars can be best
estimated from the Cyclotron Resonance Scattering Feature (CRSF) in
their X-ray spectra. The CRSFs have ever been detected from 15 X-ray
pulsars, and their surface magnetic fields are found to be distributed
within a relatively narrow range of $(1-4)\times 10^{12}$ G
(e.g.~\cite{Trumper1978}; \cite{White1983}; \cite{Mihara1995};
\cite{Makishima1999}; \cite{Coburn2002}; and references therein).

GX 304$-$1 was discovered by high-energy X-ray balloon observations
carried out since 1967 (e.g.~\cite{McClintock1971}).  It exhibits
properties typical of binary X-ray pulsars, including the large flux
variability \citep{Ricker1973}, the 272-s coherent pulsation
(\cite{Huckle1977}; \cite{McClintock1977}), and a hard X-ray
spectrum represented by a power-law with an absorption column density
$N \rm{_{H}} \sim 1 \times 10^{22}$ cm$^{-2}$ and a photon index
$\Gamma \sim$2 up to 40 keV \citep{White1983}.  A study with the Vela
5B satellite over 7 years revealed a 132.5-day periodicity of flaring
events \citep{PriedhorskyandTerrell1983}, attributable to the binary period.

  

GX 304$-$1 has been identified with a Be star system \citep{Mason1978},
showing strong shell lines (\cite{Thomas1979}; \cite{Parkes1980}) and
photometric variability \citep{Menzies1981} in the optical
wavelength.  From the visual extension ($A_V = 6.9$ mag.) to the source directions,
the distance was estimated to be $2.4 \pm 0.5$ kpc \citep{Parkes1980}.
It is consistent with the observed X-ray absorption
column density \citep{White1983}.



Since 1980, GX 304$-$1 had been in an X-ray off state
\citep{Pietsch1986} and no significant X-ray emission was detected
for 28 years.  Its quiescence was broken by the hard X-ray detection
with INTEGRAL in 2008 June \citep{Manousakis2008}.  Since then, the
source seemed to return to the active state.  
Actually, from November 2009 to January 2011, MAXI and Swift have
detected three outbursts every 132.5-day interval 
(\cite{Yamamoto2009}; \cite{Krimm2010}; \cite{Mihara2010a}).


We here report the discovery of a CRSF in RXTE and Suzaku X-ray spectra
of GX 304$-$1, obtained during the outbursts in 2010 through follow-up
observations triggered by MAXI.  We also discuss a possible change of 
the observed CRSF energy.

\section{Observations and Data Reductions}
\label{sec:obs}

\subsection{Monitoring with MAXI}
\begin{figure}
  \begin{center}
    \includegraphics[width=85mm]{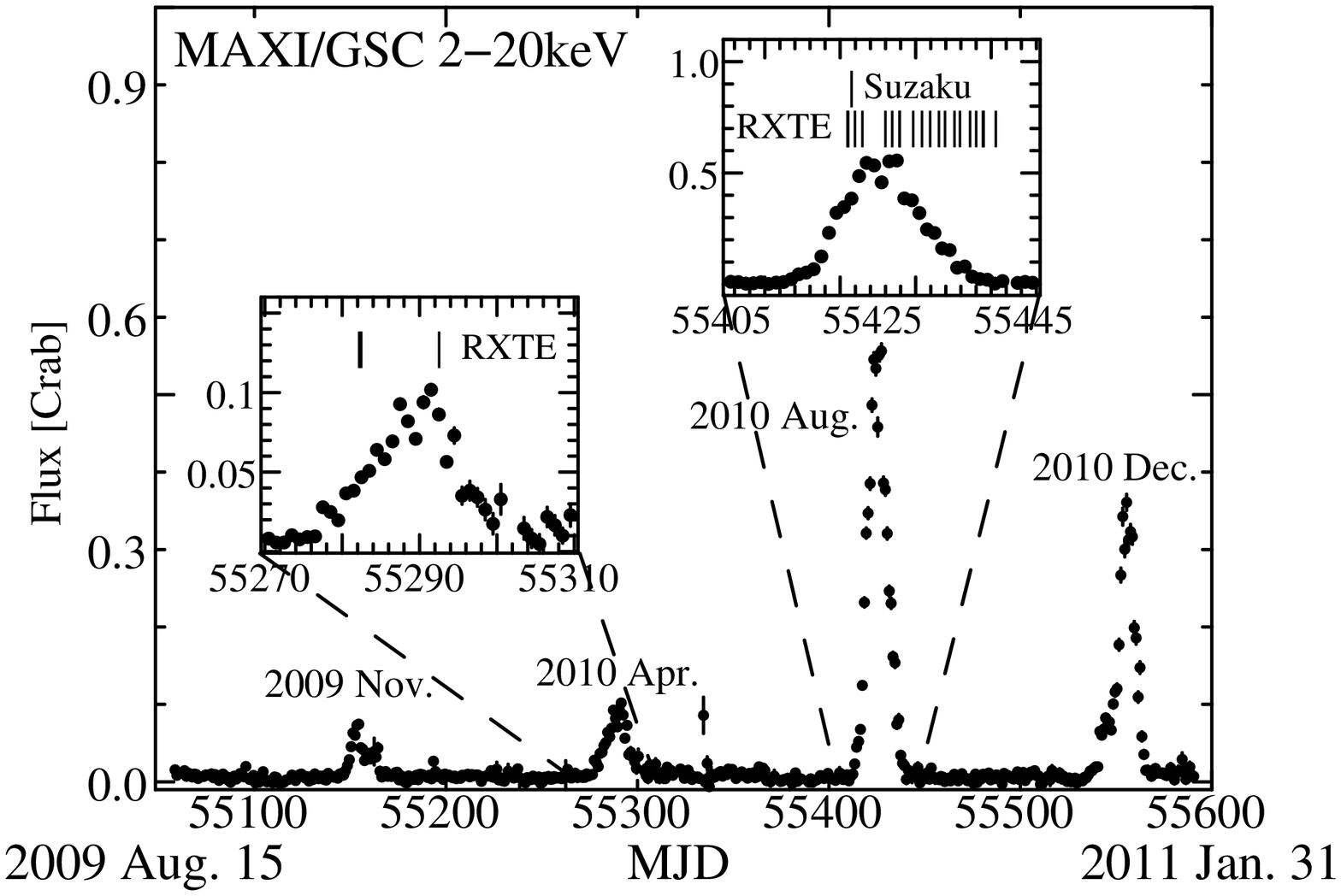}
    \caption{ MAXI/GSC light curve of GX 304$-$1 in 2--20 keV band
      from 2009 August 15 to 2011 January 31}.  The left inset
      shows a zoom up around the outburst from 2010 March 15 to April
      24, and the right inset the 
      outburst from 2010 July 28 to September 6. 
      The RXTE and Suzaku  observations
      are indicated with bars in each inset.
      
    
    \label{fig:maxi_lc}
\end{center}
\end{figure}

MAXI/GSC (\cite{matsuoka_pasj2009}, \cite{Mihara_pasj2011})
has been monitoring the flux of GX
304$-$1 since the mission start \citep{Sugizaki_pasj2011}.  
Figure \ref{fig:maxi_lc} shows the
MAXI/GSC light curve of GX 304$-$1 from 2009 August 15 (MJD=55058) to
2011 January 31 (MJD=55592).  
Four outbursts were detected with an interval of 
132.5 d, which is consistent
with the orbital period suggested from the Vela 5B data
\citep{PriedhorskyandTerrell1983}.  
They peaked on 2009 November 19 (MJD=55154), 
2010 April 1 (MJD=55287), 2010 August 15
(MJD=55423), and 2011 December 25 (MJD=55555).
%
%
The peak intensities of the first three outbursts 
gradually increased.
In the 2--20 keV band, the outburst in 2010 August reached 0.6 Crab, 
which is the highest among flaring events ever
observed from this source.
The 2010 December outburst was also bright, but did not reach the level of the 2010 August event.

\subsection{RXTE Observations}
RXTE ToO (Target of Opportunity) observations of GX 304$-$1 were
performed during the outbursts in 2010 March and August, and gave
useful data in the energy range from 3 to 250 keV with the
Proportional Counter Array (PCA: \cite{Jahoda2006}) and the
High-Energy X-ray Timing Experiment (HEXTE: \cite{Rothschild1998}).
%
%
%
The total 21 observations were carried out, 
with exposure of 0.5--5 ks each.
The observation epochs are indicated in figure
\ref{fig:maxi_lc}.



The RXTE data were reduced with the standard procedure using the relevant
analysis software in HEASOFT version 6.9 and CALDB (calibration
database) files of version 20100607, provided by NASA/GSFC
RXTE GOF (Guest Observer Facility).
PCA source spectra and background files in the 3--20 keV energy band
were extracted from the layer1 in PCU 2 alone.  
%

The hard X-ray ($>20$ keV) spectra of the source were extracted from
the HEXTE cluster-A, while backgrounds were extracted from cluster-B
and converted to cluster-A background files using the ftool {\sf
  hextebackest}.  Since the HEXTE background spectra reproduced by the
standard method are known to have a relatively large calibration
uncertainty at around 63 keV for the data after 
2009 December\footnote[1]{http://heasarc.gsfc.nasa.gov/docs/xte/xhp\_new.html}, we
chose, for the subsequent spectral analysis, observations whose 
signal-to-background ratio is higher than 30\% at 50 keV.  
Table \ref{tab:ObsRXTE} summarizes the log of the selected twelve
observations.

%

\subsection{Suzaku Observation}

A Suzaku ToO observation of GX 304$-$1 was performed on 2010 August 13,
two days before the outburst maximum.
It was triggered by the MAXI detection of
the rapid flux increase \citep{Mihara2010a}.
The Suzaku data covers an energy band from 0.5 to 500 keV,
using the X-ray Imaging Spectrometer (XIS: \cite{Koyama2007}) and the Hard
X-ray Detector (HXD: \cite{Takahashi2007}, \cite{Kokubun2007}).  The
target was placed at the HXD nominal position on the detectors.  The
XIS was operated in the normal mode with 1/4-window and 0.5 s burst
options, which gives a time resolution of 2 s.  The HXD was
operated in the nominal mode.
Table \ref{tab:ObsSuzaku} summarizes the observation log.

The data reduction and analysis were performed with the standard
procedure using the Suzaku analysis software in HEASOFT version 6.9
and the CALDB files version 20100812, provided by NASA/GSFC Suzaku
GOF.  All obtained data were first reprocessed by {\sf aepipeline} to
utilize the latest calibration.  The net exposures after the
standard event-screening process were 5.1 ks with the XIS and 12.1 ks
with the HXD. The former is significantly shorter than the latter because
of the 0.5 s burst option.
%
%
%
%
The background spectra 
for HXD-PIN and HXD-GSO were created with the standard
manner, using the archived background event files provided via the
Suzaku GOF.
This process also removes the Cosmic X-ray Background (CXB) from
the HXD-PIN data, while that in the HXD-GSO data is negligible \citep{Fukazawa2009}.
%
%
%
%
%
%
After subtracting the backgrounds, the source was detected
significantly at an intensity of $36.3\pm 0.05$ counts s$^{-1}$ with
PIN in 15--75 keV, and $2.46\pm 0.05$ counts s$^{-1}$ with GSO in
50--130 keV.

\begin{table*}
\begin{center}
  \caption{Log of RXTE Observations of GX 304$-$1 in the 2010 August Outburst}
  \label{tab:ObsRXTE}
  \begin{tabular}{lccccp{1pt}cc}
  \hline              
  \hline              
  Date     & Obs ID & Obs Time & \multicolumn{2}{c}{PCA (3--20 keV)$^{\dagger }$} & & \multicolumn{2}{c}{HEXTE (20--100 keV)} \\ 
  \cline{4-5} \cline{7-8}
  (2010   &  (95417-01-)  & Start / End  & Exp.          & Rate                     & & Exp.         & Rate                    \\ 
   Aug.)   &             &  (UT)     & (ks)          & (counts s$^{-1}$)        & & (ks)         & (counts s$^{-1}$)       \\ 
  \hline
   13a   & 03-03       & 03:32 / 04:20    & 2.3           & 941.3$ \pm$1.1        & & 1.4          & 147.6$ \pm$0.4       \\
   13b   & 03-00       & 04:44 / 06:37    & 3.7           & 997.9$ \pm$1.1        & & 2.3          & 155.0$ \pm$0.3       \\
   14    & 03-01       & 01:37 / 04:35    & 5.4           & 1060.0$ \pm$1.1       & & 1.5          & 163.9$ \pm$0.4       \\
   15    & 03-02       & 01:59 / 04:45    & 6.1           & 1143.0$ \pm$1.2       & & 2.0          & 175.4$ \pm$0.3       \\
   18    & 04-00       & 02:25 / 03:57    & 3.3           & 1130.0$ \pm$1.3       & & 2.1          & 163.9$ \pm$ 0.3       \\
   19    & 04-01       & 01:57 / 02:57    & 3.2           & 1211.0$ \pm$1.4       & & 2.0          & 176.3$ \pm$0.4       \\
   20    & 05-00       & 00:02 / 01:00    & 3.2           & 1110.0$ \pm$1.3       & & 1.9          & 159.6$ \pm$0.3       \\
   21    & 05-01       & 20:33 / 20:55    & 1.0           & 774.4$ \pm$1.2        & & 0.6          & 101.6$ \pm$0.5       \\
   22    & 05-02       & 23:58 / 00:43    & 2.0           & 654.8$ \pm$0.9        & & 1.2          & 82.9$ \pm$0.4        \\
   24    & 05-03       & 02:40 / 03:44    & 3.4           & 546.7$ \pm$0.7        & & 2.1          & 64.1$ \pm$0.2        \\
   25    & 05-04       & 05:44 / 06:12    & 1.2           & 422.2$ \pm$0.7        & & 0.9          & 48.1$ \pm$0.4        \\
   26    & 05-05       & 00:42 / 01:16    & 1.4           & 376.5$ \pm$0.7        & & 0.8          & 44.6$ \pm$0.4        \\
\hline
\end{tabular}
\end{center}
\hspace*{2cm} 
\begin{tabular}{l}
$^{\dagger}$ PCU2 only. \\
\end{tabular}
\end{table*}


\begin{table*}
\begin{center}
  \caption{Log of Suzaku Observation of GX 304--1 in the 2010 August Outburst}
  \label{tab:ObsSuzaku}
  \begin{tabular}{ccccp{1pt}ccp{1pt}cc}
    \hline              
    \hline              
    Date        & Obs Time  & \multicolumn{2}{c}{XIS0 (1--10 keV)}    & & \multicolumn{2}{c}{HXD-PIN (15--75 keV)} & & \multicolumn{2}{c}{HXD-GSO (50--130 keV)} \\
    \cline{3-4}\cline{6-7}\cline{9-10}
    (2010     & Start/End      & Exp.          & Rate                   & & Exp.           & Rate                   & & Exp.           & Rate                   \\
     Aug.)     &   (UT)      & (ks)          & (counts s$^{-1}$)      & & (ks)           & (counts s$^{-1}$)      & & (ks)           & (counts s$^{-1}$)      \\
    \hline
    13        & 16:19/23:00    & 5.13          & 150.6$\pm$0.2         & & 12.14           & 36.25$\pm$0.05       & & 12.14          & 2.56$\pm$0.05         \\
    \hline
  \end{tabular}
\end{center}
\hspace*{1cm} Observation ID $=$ 905002010
\end{table*}

\section{Analysis and Results}
\label{sec:results}

The barycentric pulsation period was derived 
to be 275.46 s during the Suzaku observation, from the folding analysis of the HXD-PIN data.

The RXTE and Suzaku observations both provide us with an opportunity
to search for CRSFs that have not been detected from GX 304$-$1 in the X-ray energy
band up to 40 keV \citep{White1983}.  
Hereafter we concentrate on the
analysis of pulse-phase-averaged spectra for CRSFs. 

We present results using the data of 
the PCA (3--20 keV) and the HEXTE
(20--100 keV) from RXTE,
and those of
HXD-PIN (15--75 keV) and HXD-GSO (50--130 keV) from Suzaku.
The
Suzaku XIS data were not used in the present paper, because they suffer
considerably from event pile-up.
%
All the spectral fits were carried out on XSPEC version 12.6.0.


\subsection{CRSF in X-ray Spectra by RXTE and Suzaku}
\label{sec:suzaku_rxte_spec_fit}

\begin{figure}
  \begin{center}
    \includegraphics[width=85mm]{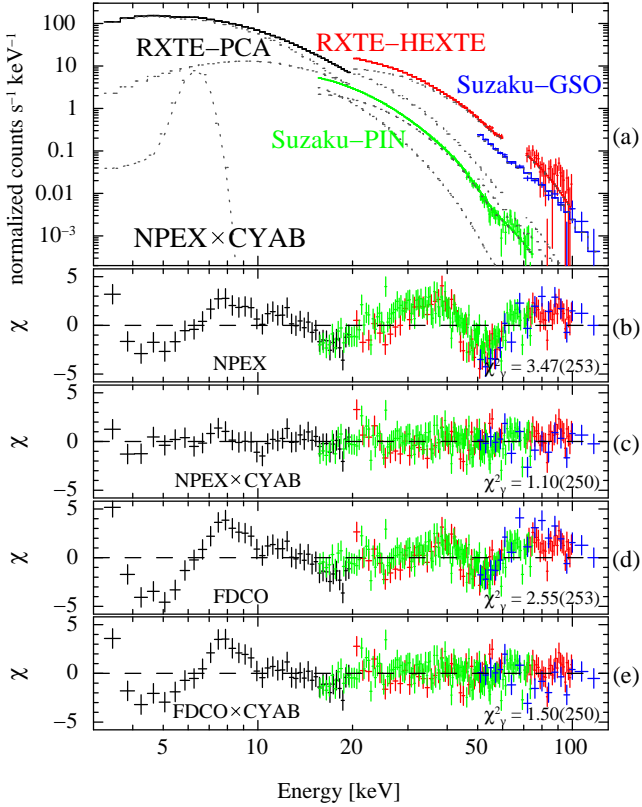}
  \end{center}
  \caption{ X-ray spectra of GX 304$-$1 observed by
    RXTE and Suzaku on August 13-14.  
    (a) Data and the best-fit spectral models of NPEX$\times$CYAB.
    (b)-(e) Residuals from the best-fit NPEX, NPEX$\times$CYAB, FDCO, and FDCO$\times$CYAB
    models, respectively.
  }
 \label{fig:spec_xte_suzaku_fit}
\end{figure}

We first performed joint spectral fits to the data taken by RXTE and
Suzaku during 12 hours from August 13 16:00 (UT), as presented in
figure \ref{fig:spec_xte_suzaku_fit}.  Since these observations are
not exactly simultaneous, the average flux can be different between the two data sets. 
We thus introduced a parameter representing
relative normalization of the over all model, and allowed it to take defferent
values among the PCA, HEXTE, HXD-PIN, and HXD-GSO spectra.
The four values of this parameter agreed with one another within calibration uncertainties.

We here examined the validity of the RXTE-HEXTE background spectrum.
The energy band from 61 keV to 71 keV was ignored in all the
subsequent analysis since artificial structures are known to remain
for the data taken after 
2009 December.
We also attempted to change
the background scale factor and checked if any artificial features
remain in the residual.  Assuming that there is no significant source
flux above the background in a higher energy band of 150--250 keV, the
best background scale factor was obtained to be 1.1.  We employed this
value when subtracting the HEXTE background.  The validity was
further confirmed from the consistency with the Suzaku data.

We employed 
a cutoff power-law (cutoffpl model in XSPEC),
an NPEX (Negative and Positive power laws with
exponential cutoff: \cite{Mihara1995}; \cite{Makishima1999}) 
or an FDCO (Fermi-Dirac cutoff power-law: \cite{Makishima1999}) 
model to reproduce the 
continuum from 3 keV to 130 keV.
The cutoffpl model was far from successful, with reduced chi-squared
$\chi^{2}_{\nu}=16.8$ for degrees of freedom $\nu=254$.
Thus it is excluded in the spectral analysis hereafter.
In the NPEX model
we left free all parameters but one :
the positive power-law index, $\alpha_{2}$, was fixed at 2.0, representing
a Wien peak, because it was not well constrained by the data.  The
fit with either NPEX or FDCO model alone was 
unacceptable ($\chi^{2}_{\nu}=3.47$ for $\nu=253$, and 
$\chi^{2}_{\nu}=2.55$ for $\nu=253$, respectively).
As shown in figure \ref{fig:spec_xte_suzaku_fit} (b) and (d),
the residuals similarly exhibit absorption
features around 20--30 keV and 40--60 keV in 
both the RXTE and the Suzaku spectra respectively.

We then multiplied the continum models with cyclotron absorption (CYAB) factors
(\cite{Mihara1990}; \cite{Makishima1999}).  
The NPEX model with a single
CYAB feature was accepted within the 90\%
confidence limit ($\chi^{2}_{\nu}=1.10$ for $\nu=250$)
as shown in figure
\ref{fig:spec_xte_suzaku_fit} (c).
The fundamental resonance energy was obtained to be $E_{\rm a} = 53.7^{+0.7}_{-0.6}$ keV.
In contrast, the FDCO model with a CYAB was not acceptable
($\chi^{2}_{\nu}=1.50$ for $\nu=250$),
leaving wavy residuals in 3--10 keV in figure
\ref{fig:spec_xte_suzaku_fit} (e).

The NPEX model
with two CYAB features that represent the fundamental harmonics $E_{\rm a1}$
$\sim$20 keV, and the second harmonics $E_{\rm a2} =$ 2$E_{\rm a1}$ was also examined.
However, the fit was not improved at all
($\chi^{2}_{\nu}=1.11$ for $\nu=248$) and the depth of the 
fundamental harmonic was zero within the statistical error.
Therefore, both the
RXTE and the Suzaku data confirm the presence of a fundamental CRSF at
about $E_{\rm a}=$ 54 keV, and imply that the NPEX continum is most successful 
among the three models tested. Table \ref{tab:suzaku_xte_fit_param} summarizes these
fitting results and the best-fit model parameters.
As given there, the FDCO model
(though not acceptable) gives a consistent resonance energy.

%

%
%



\begin{table*}
  \begin{center}
    \caption{Summary of joint fits to Suzaku and RXTE spectra taken on 2010 August 13-14}
    \label{tab:suzaku_xte_fit_param}
    \begin{tabular}{lcccccc}
      \hline
      \hline
      Parameter                               & \multicolumn{6}{c}{Model}                                                                                                         \\
                                              & cutoffpl     & FDCO           & FDCO$\times$CYAB         & NPEX           & NPEX$\times$CYAB         & NPEX$\times$CYAB2$^{a}$        \\
      \hline
      $N_{\rm H}$ ($10^{22}$ cm$^{-2}$)       & $ 0.00 $       & $ 5.93 $       & $ 5.26_{-0.24}^{+0.23} $ & $ 4.22 $       & $ 3.13_{-0.26}^{+0.24} $ & $ 3.08_{-0.23}^{+0.33} $ \\
      $I_{\rm Fe}$$^{b}$ ($\times 10^{-2}$)   & $ 1.90 $       & $ 0.67 $       & $ 0.82_{-0.13}^{+0.13} $ & $ 0.81 $       & $ 0.91_{-0.13}^{+0.13} $ & $ 0.91_{-0.14}^{+0.13} $ \\
      $A_{1}$$^{c}$ ($\times 10^{0}$)         & $ 0.43 $       & $ 1.73 $       & $ 1.60_{-0.05}^{+0.05} $ & $ 0.92 $       & $ 0.72_{-0.03}^{+0.03} $ & $ 0.71_{-0.04}^{+0.03} $ \\
      $\alpha_1$                              & $ 0.35 $       & $ 1.33 $       & $ 1.25_{-0.02}^{+0.02} $ & $ 0.57 $       & $ 0.49_{-0.02}^{+0.02} $ & $ 0.50_{-0.02}^{+0.02} $ \\
      $E_{\rm cut}$ (keV)                     & ---            & $ 31.7 $       & $ 27.7_{-1.1}^{+0.9}   $ & ---            & ---                      & ---                      \\
      $kT/E_{\rm fold}$ (keV)                 & $11.2$         & $ 9.0  $       & $ 11.8_{-0.5}^{+0.7}   $ & $ 6.5  $       & $ 7.4_{-0.2}^{+0.2}    $ & $ 7.5_{-0.2}^{+0.1}    $ \\
      $A_{2}$$^{c}$ ($\times 10^{-4}$)        & ---            & ---            & ---                      & $ 9.4  $       & $ 5.2_{-0.6}^{+0.5}    $ & $ 5.1_{-0.8}^{+0.8}    $ \\
      $E_{\rm a1}$ (keV)                      & ---            & ---            & $ 54.5_{-0.9}^{+1.1}   $ & ---            & $ 53.7_{-0.6}^{+0.7}   $ & $ 26.9_{-0.3}^{+0.3}   $ \\
      $W_{1}$ (keV)                           & ---            & ---            & $ 9.8_{-2.2}^{+2.9}    $ & ---            & $ 10.2_{-2.0}^{+2.3}   $ & $ 1.0_{-1.0}           $ \\
      $D_{1}$                                 & ---            & ---            & $ 0.75_{-0.09}^{+0.13} $ & ---            & $ 0.73_{-0.06}^{+0.09} $ & $ 0.01_{-0.01}^{+0.02} $ \\
      $W_{2}$ (keV)                           & ---            & ---            & ---                      & ---            & ---                      & $ 10.9_{-2.4}^{+2.1}   $ \\
      $D_{2}$                                 & ---            & ---            & ---                      & ---            & ---                      & $ 0.75_{-0.08}^{+0.15} $ \\
      $\chi^2_{\nu}$ $(\nu)$                  & $16.8$ $(254)$ & $2.55$ $(253)$ & $1.50$ $(250)$           & $3.47$ $(253)$ & $1.10$ $(250)$           & $ 1.11$ $(248)$          \\

\hline
    \end{tabular}

  \end{center}

  \begin{tabular}{l}
    All errors represent the 90\% confidence limits of the statistical uncertainties.\\
    $^{a}$ CYAB2: $E_{\rm a2}$ energy is fixed to $2 E_{\rm a1}$.\\
    $^{b}$ units in photons s$^{-1}$ cm$^{-2}$.\\
    $^{c}$ units in photons s$^{-1}$ cm$^{-2}$ keV$^{-1}$ at 1 keV.\\

  \end{tabular}
    
\end{table*}

\subsection{CRSF energy variation}

As shown in figure \ref{fig:maxi_lc}, the RXTE observations in 2010
August covered the peak-to-descent phase of the outburst on an almost
daily basis.  The data enable us to investigate spectral variations
in this period.

With the same procedure as described in subsection
\ref{sec:suzaku_rxte_spec_fit}, model fits to individual spectra taken
in these RXTE observations and the Suzaku were performed.  
By artificially changing the HEXTE background by $\pm$5\% of the nominal value,
we confirmed that the obtained best-fit parameters are not sensitive to
the background uncertainty.

\begin{figure}
  \begin{center}
    \includegraphics[width=85mm]{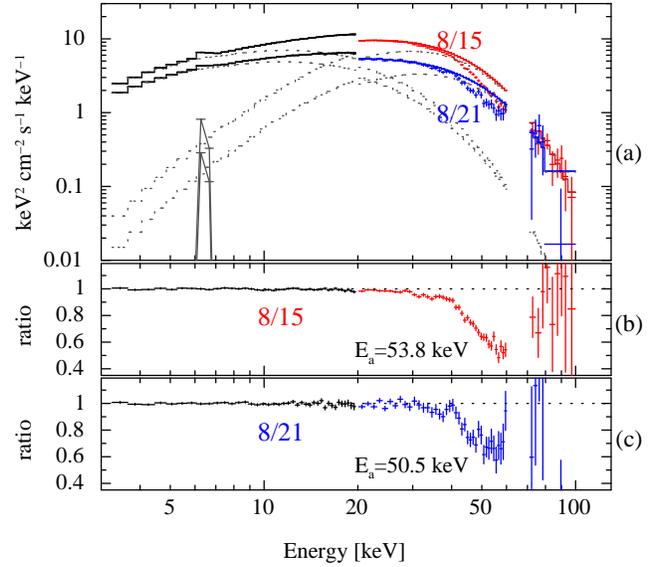}
  \end{center}
  \caption{
    Comparison of X-ray spectra taken by RXTE on August 15 and 21.
    (a) Unfolded spectra and best-fit NPEX$\times$CYAB models.
    The negative and positive power-law components are shown in the dotted lines.
    (b) Data-to-model ratio for the August 15 spectrum,shown after removing the CYAB factor from the best-fit NPEX$\times$CYAB fit.
    (c) The same as (b) but for the August 21 spectrum.
  }
 \label{fig:spec_nufnu_ratio}
\end{figure}

These spectral fits with NPEX model revealed that the CYAB feature is
required by all the spectra of the selected observations with a
significance above 90\%.  The obtained best-fit parameters are
summarized in table \ref{tab:NPEXCYAB}, where the CRSF energy is seen
to vary, beyond the fitting errors, by $\sim$6\% among the
observations.  Figure \ref{fig:spec_nufnu_ratio} illustrates the
difference of the CRSF feature in the spectra taken on August 15 and 21. 
Thus, the resonance energy appears to have really changed between the 
two data sets.

\begin{figure}
  \begin{center}
    \FigureFile(85mm,){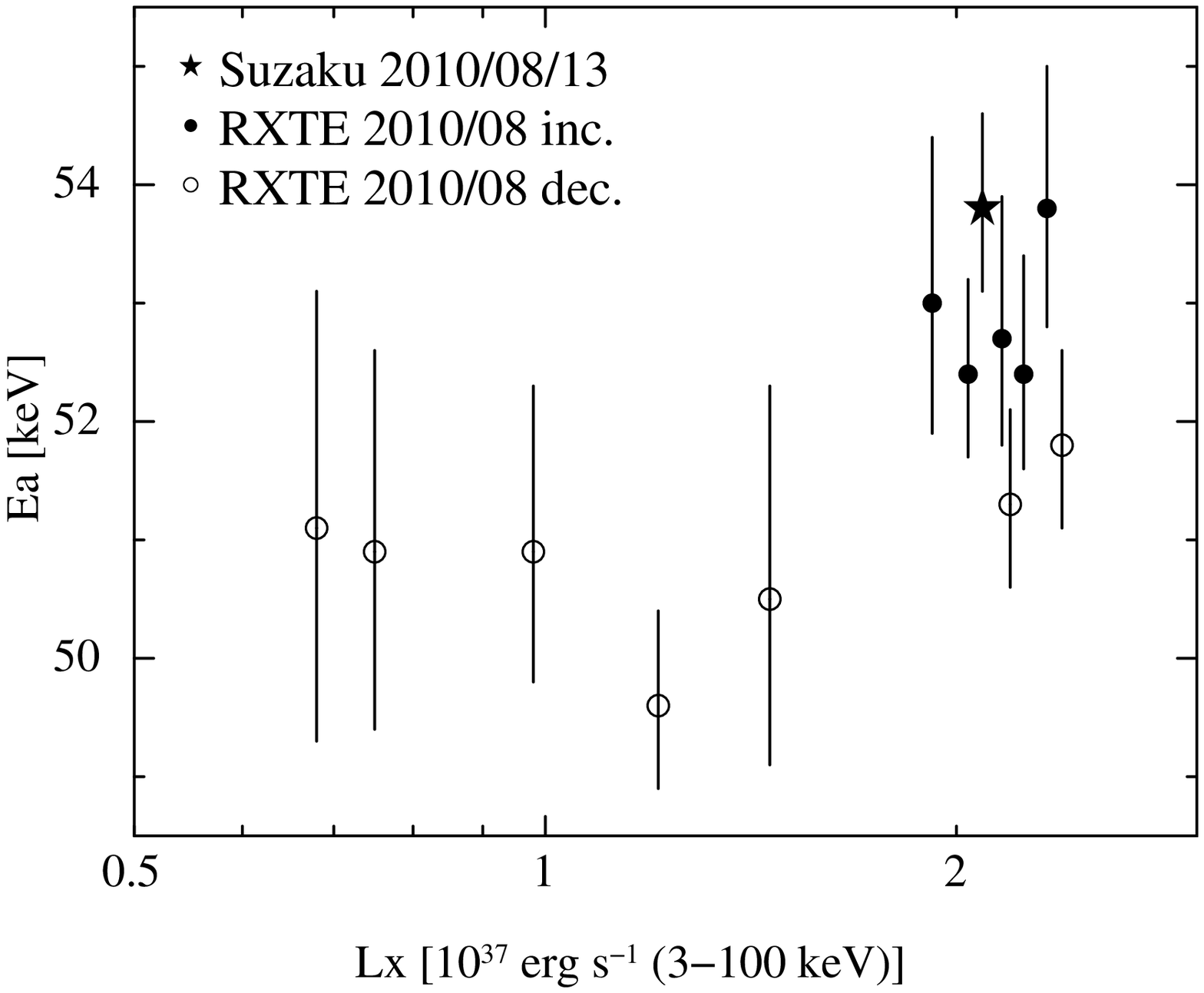}
  \end{center}
  \caption{ 
    Relation between the CRSF energy and the 3-100 keV X-ray
    luminosity during the 2010 August outburst.
    Data points obtained from 
    the RXTE observations of increasing and decreasing phases, and 
    the Suzaku observation are marked with
    filled circles, open circles, and a star}, respectively.  The vertical error bars
    represent the 90\% confidence limits of the statistical
    uncertainty, obtained from the model fits.  
  
\label{fig:ea_lx_relation}
\end{figure}

Figure \ref{fig:ea_lx_relation}
plots the relation between the the CRSF energy and
the 3--100 keV luminosity, estimated from the best-fit
spectral models.  
The results allow at least two alternative interpretations. 
  One is that the the CRSF energy depends positively on the X-ray luminosity.
  The other is that the CRSF energy  splits into two regimes, $\sim$50 keV 
  and $\sim$54 keV, depending possibly on the outburst phase
  (e.g. \cite{Caballero2008}).

\begin{table*} 
\begin{center}
\caption{Best-fit parameters of the NPEX$\times$CYAB models to spectra by RXTE and Suzaku in 2010 August outburst}
\label{tab:NPEXCYAB}
\small
\begin{tabular}{lcccccccccccc}
\hline
\hline
Date   & $N_{\rm H}$              & $I_{\rm Fe}^{a}$         & $A_{1}$$^{b}$            & $\alpha_1$               & $A_{2}$$^{b}$          & $kT$                     & $E_{\rm a}$            & $W_{1}$        & $D_{1}$  & $\chi^2_{\nu}$ $(\nu)$& $L_{\rm x}^{c}$\\
\multicolumn{2}{r}{ ($10^{22}$cm$^{-2}$) }    & ($\times 10^{-2}$)       & ($\times 10^{0}$)        &                          & ($\times 10^{-5}$)     & (keV)                    & (keV)                  & (keV)          &          &                       &                \\
\hline
13a      & $ 3.00_{-0.31}^{+0.30} $ & $ 1.01_{-0.13}^{+0.14} $ & $ 0.69_{-0.04}^{+0.04} $ & $ 0.49_{-0.03}^{+0.02} $ & $ 56.5_{-7.8}^{+6.3} $ & $ 7.2_{-0.2}^{+0.3} $ & $ 53.0_{-1.1}^{+1.4} $ & $ 7.4_{-2.4}^{+3.5} $ & $ 0.67_{-0.08}^{+0.09} $ & 1.10 (98) & 1.92 \\
13b      & $ 2.95_{-0.29}^{+0.27} $ & $ 0.96_{-0.13}^{+0.14} $ & $ 0.71_{-0.06}^{+0.04} $ & $ 0.47_{-0.02}^{+0.02} $ & $ 58.6_{-6.3}^{+5.3} $ & $ 7.2_{-0.1}^{+0.2} $ & $ 52.4_{-0.7}^{+0.8} $ & $ 6.5_{-1.7}^{+2.3} $ & $ 0.65_{-0.06}^{+0.07} $ & 1.13 (98) & 2.04 \\
13$^{\dagger}$  &---               &---                       & $ 0.53_{-0.20}^{+0.42} $ & $ 0.38_{-0.20}^{+0.27} $ & $ 57.6_{-11.8}^{+19.9} $ & $ 7.2_{-0.3}^{+0.3} $ & $ 53.8_{-0.7}^{+0.8} $ & $ 7.4_{-2.0}^{+2.4} $ & $ 0.75_{-0.08}^{+0.09} $ & 1.05 (146)& 2.09 \\
14       & $ 2.94_{-0.28}^{+0.27} $ & $ 1.00_{-0.14}^{+0.14} $ & $ 0.74_{-0.04}^{+0.04} $ & $ 0.46_{-0.02}^{+0.02} $ & $ 58.7_{-8.8}^{+6.7} $ & $ 7.3_{-0.2}^{+0.3} $ & $ 52.7_{-0.9}^{+1.2} $ & $ 8.1_{-2.5}^{+3.3} $ & $ 0.60_{-0.07}^{+0.09} $ & 0.96 (98) & 2.16 \\
15       & $ 2.99_{-0.27}^{+0.26} $ & $ 1.20_{-0.14}^{+0.14} $ & $ 0.78_{-0.03}^{+0.04} $ & $ 0.46_{-0.02}^{+0.02} $ & $ 60.8_{-8.7}^{+6.6} $ & $ 7.4_{-0.2}^{+0.3} $ & $ 53.8_{-1.0}^{+1.2} $ & $ 9.6_{-2.5}^{+3.4} $ & $ 0.66_{-0.07}^{+0.12} $ & 0.97 (98) & 2.33 \\
18$^{d}$ & $ 2.77_{-0.27}^{+0.26} $ & $ 1.10_{-0.15}^{+0.14} $ & $ 0.79_{-0.04}^{+0.04} $ & $ 0.47_{-0.02}^{+0.02} $ & $ 45.2_{-7.0}^{+5.8} $ & $ 7.6_{-0.2}^{+0.3} $ & $ 52.4_{-0.8}^{+1.0} $ & $ 7.7_{-2.1}^{+2.9} $ & $ 0.60_{-0.06}^{+0.09} $ & 0.86 (56) & 2.24 \\
19       & $ 2.32_{-0.26}^{+0.26} $ & $ 1.33_{-0.16}^{+0.15} $ & $ 0.77_{-0.04}^{+0.04} $ & $ 0.43_{-0.02}^{+0.02} $ & $ 42.7_{-8.0}^{+6.8} $ & $ 7.7_{-0.2}^{+0.4} $ & $ 51.8_{-0.7}^{+0.8} $ & $ 8.1_{-2.2}^{+2.9} $ & $ 0.56_{-0.06}^{+0.10} $ & 0.88 (98) & 2.39 \\
20       & $ 2.83_{-0.27}^{+0.25} $ & $ 1.14_{-0.15}^{+0.14} $ & $ 0.77_{-0.04}^{+0.04} $ & $ 0.47_{-0.02}^{+0.02} $ & $ 39.0_{-5.8}^{+5.1} $ & $ 7.7_{-0.2}^{+0.3} $ & $ 51.3_{-0.7}^{+0.8} $ & $ 6.0_{-1.9}^{+2.4} $ & $ 0.48_{-0.06}^{+0.06} $ & 1.19 (98) & 2.19 \\
21       & $ 3.13_{-0.34}^{+0.35} $ & $ 0.42_{-0.13}^{+0.13} $ & $ 0.69_{-0.04}^{+0.04} $ & $ 0.59_{-0.03}^{+0.03} $ & $ 22.5_{-5.8}^{+5.6} $ & $ 7.9_{-0.4}^{+0.5} $ & $ 50.5_{-1.4}^{+1.8} $ & $ 6.0_{-3.4}^{+4.5} $ & $ 0.52_{-0.13}^{+0.16} $ & 0.85 (98) & 1.46 \\
22       & $ 3.06_{-0.29}^{+0.29} $ & $ 0.56_{-0.10}^{+0.10} $ & $ 0.67_{-0.03}^{+0.03} $ & $ 0.68_{-0.02}^{+0.03} $ & $ 14.2_{-2.9}^{+3.1} $ & $ 8.3_{-0.2}^{+0.4} $ & $ 49.6_{-0.7}^{+0.8} $ & $ 4.3_{-1.9}^{+2.2} $ & $ 0.79_{-0.16}^{+0.24} $ & 0.97 (98) & 1.21 \\
24       & $ 3.39_{-0.26}^{+0.26} $ & $ 0.31_{-0.08}^{+0.08} $ & $ 0.62_{-0.03}^{+0.03} $ & $ 0.74_{-0.03}^{+0.03} $ & $  9.7_{-1.9}^{+1.9} $ & $ 8.5_{-0.1}^{+0.3} $ & $ 50.9_{-1.1}^{+1.4} $ & $ 6.3_{-2.2}^{+3.0} $ & $ 0.53_{-0.10}^{+0.11} $ & 1.14 (98) & 0.98 \\
25       & $ 4.05_{-0.34}^{+0.36} $ & $ 0.23_{-0.08}^{+0.09} $ & $ 0.50_{-0.03}^{+0.03} $ & $ 0.76_{-0.04}^{+0.04} $ & $  5.7_{-1.6}^{+1.9} $ & $ 8.7_{-0.1}^{+0.4} $ & $ 50.9_{-1.5}^{+1.7} $ & $ 6.0$ fix$^{e} $     & $ 0.70_{-0.19}^{+0.21} $ & 1.43 (99) & 0.75 \\
26       & $ 4.49_{-0.34}^{+0.35} $ & $ 0.23_{-0.08}^{+0.08} $ & $ 0.49_{-0.02}^{+0.04} $ & $ 0.81_{-0.04}^{+0.05} $ & $  3.9_{-1.3}^{+1.5} $ & $ 9.1_{-0.2}^{+0.8} $ & $ 50.4_{-1.5}^{+1.8} $ & $ 2.7_{-2.7}^{+3.3} $ & $ 0.87_{-0.37}^{+1.87} $ & 1.26 (98) & 0.63 \\
26       & $ 4.59_{-0.35}^{+0.36} $ & $ 0.23_{-0.08}^{+0.08} $ & $ 0.50_{-0.03}^{+0.03} $ & $ 0.84_{-0.04}^{+0.05} $ & $  3.6_{-1.1}^{+1.3} $ & $ 9.4_{-0.8}^{+0.7} $ & $ 51.1_{-1.8}^{+2.0} $ & $ 6.0$ fix$^{e} $     & $ 0.62_{-0.19}^{+0.18} $ & 1.27 (99) & 0.68 \\
  \hline
\end{tabular}
\end{center}
\begin{tabular}{l}
  All errors represent the 90\% confidence limits of the statistical uncertainties.\\
  $^{a}$ units in photons s$^{-1}$ cm$^{-2}$.\\
  $^{b}$ units in photons s$^{-1}$ cm$^{-2}$ keV$^{-1}$ at 1 keV.\\
  $^{c}$ X-ray luminosity in 3--100 keV in units of 10$^{37}$ ergs s$^{-1}$.\\
  $^{d}$ HEXTE standard data are used. (HEXTE science data are used for other days).\\
  $^{e}$ The width is fixed.\\
  $^{\dagger}$ Suzaku data. All others are from RXTE data.\\
\end{tabular}
\end{table*}

\section{Discussion}


We analyzed the broadband X-ray (3--130 keV) spectra of GX 304$-$1 
obtained by RXTE and Suzaku, in ToO observations covering the two
outbursts in 2010 detected by MAXI.
A signature of CRSF was discovered at 54
keV from both the RXTE and the Suzaku data taken on August 13.  It is
the first detection of the CRSF from this source \citep{Mihara2010b}.
\citet{Sakamoto2010} reported a Swift-BAT confirmation of the
CRSF at around 50 keV from the spectrum accumulating data from August
12 to 17.

The CRSF energy of 54 keV exceeds that of A 0535$+$26 ($\sim$45 keV:
\cite{Terada2006}), and becomes the highest among the X-ray binary
pulsars whose CRSF parameters are well determined.  The surface
magnetic field strength is estimated to be $4.7 \times 10^{12}$
$(1+z_g)$ G, where $z_g$ represents the gravitational redshift.
\citet{Makishima1999} examined the relation between the magnetic field
strength estimated from the CRSF and the pulsation period in X-ray
binary pulsars, and discussed a group of ``slow rotators'';
represented by such sources as Vela X-1 and GX 301$-$2, these objects
have much longer pulsation periods than would be expected if they were
in rotational equilibria.  The obtained field strength of $4.7 \times
10^{12}$ G, and the pulsation period of 275.46 s measured during the Suzaku observation, place GX 304$-$1 just in
the range of the typical slow rotators.



We performed spectral analysis of the RXTE data covering the outburst
in 2010 August on an almost daily basis.  The CRSF has also been
confirmed in 10 RXTE observations in which the source was bright
enough.  Therefore, the CRSF is a persistent effect of this object.
However, the CRSF energy was observed to 
vary, either in a positive correlation with the luminosity,
  or in a bimodal manner with $E_{\rm a}\sim$50 keV and $E_{\rm a}\sim$54 keV.

Variations of the CRSF energy during a single outburst have been
observed from 4U~0115$+$63 (\cite{Mihara1998}, \cite{Mihara2004},
\cite{Nakajima2006}) and X~0331$+$53 (V~0332$+$53) (\cite{Mowlavi2006};
\cite{Tsygankov2006}; \cite{Nakajima2010}).  
However, in contrast to the behavior of GX 304$-$1 revealed in the
present study, these two objects show negative 
correlations, that the CRSF energy decreases as the 
luminosity increases.
The negative correlation can be explained by presuming that the
cyclotron-scattering photosphere gets higher when the accretion rate
increased in the super-Eddington accretion regime \citep{Mihara1998}.


A positive correlation between the CRSF energy and the luminosity
has been seen in the long-term behavior of Her X-1 over
multiple outbursts (\cite{Gruber2001}; \cite{Staubert2007}).
In  additions, 
different CRSF energies were measured between two orbital phases in GX 301$-$2
\citep{LaBarbera2005}.  The behavior is expected in a sub-Eddington
accretion, where the cyclotron-scattering photosphere is lowered by
the dynamical pressure of the accretion \citep{Staubert2007}.  The
observed behavior of GX 304$-$1, if interpreted as showing a positive
dependence of $E_{\rm a}$ on the luminosity,
may be a manifestation of the same effects, and regarded as 
the first example that the relation was observed in a
single outburst.  Indeed, the fraction of the CRSF-energy change,
$\Delta E_{\rm a}/E_{\rm a}\sim$ 6\%, is similar to that observed in
Her X-1, and reasonably agrees with that of the quantitative estimate
in these situations in \citet{Staubert2007}.

A bimodal change in the CRSF energy was observed from A 0535$+$26
by \citet{Caballero2008}; they  measured the resonance energy at $\sim$46 keV
in the 2005 outburst, and at $\sim$54 keV during its pre-putburst, even though
the luminosity was comparable on the two occasions. 
\citet{Postnov2008} interpreted this effect in terms of magnetospheric instabilities
between the accretion disk and the neutron-star magnetosphere at the
onset of accretion.  The same scenario may apply also to our figure 4,
if it is interpreted as representing two typical values of $E_{\rm a}$.




Since the mission started on 2010 August 15,
MAXI detected four X-ray outbursts from GX 304$-$1 by 132.5-day
intervals. As reported by \citet{Manousakis2008}, this confirmed
the recurrence of the source activities after 28 years of X-ray disappearance.
The source may have returned to the
active state such that it had been in until 1980.  We urge 
continuous monitoring of this source, and follow-up observations of
outbursts with hard X-ray instruments for further studies of the
CRSF behaviors.

\bigskip


This research was partially supported by the Ministry of Education,
Culture, Sports, Science and Technology (MEXT), Grant-in-Aid for
Science Research (A) 20244015 and for Young Scientists (B) 21740140.


\end{document}